\RequirePackage{commath, amsmath, amssymb, amsfonts}
\documentclass[10pt, a4paper]{iopart}
\usepackage{xcolor}
\usepackage{physics}
\usepackage{graphicx}
\usepackage{amsmath}
\usepackage{tikz}
\usepackage{hyperref}
\usepackage{physics}
\usepackage{amsmath}
\usepackage{mathdots}
\usepackage{yhmath}
\usepackage{cancel}
\usepackage{color}
\usepackage{multirow}
\usepackage{amssymb}
\usepackage{tabularx}
\usepackage{booktabs}
\usepackage{makecell}

\begin{document}
\title{Absorption \textit{versus} Adsorption:\\High-Throughput Computation of Impurities in 2D Materials}
\author{Joel Davidsson$^{1,*,\dagger}$, Fabian Bertoldo$^{2,*}$, Kristian S. Thygesen$^2$, and Rickard Armiento$^1$}
\address{$^1$Department of Physics, Chemistry and Biology, Link\"oping University, SE-581 83 Link\"oping, Sweden}
\address{$^2$CAMD, Computational Atomic-Scale Materials Design, Department of Physics, Technical University of Denmark, 2800 Kgs. Lyngby, Denmark}

\address{$^*$These authors contributed equally: Joel Davidsson, Fabian Bertoldo}
\address{$^\dagger$Corresponding author: joel.davidsson@liu.se}
\vspace{10pt}

\begin{abstract}
Doping of a two-dimensional (2D) material by impurity atoms occurs \textit{via} two distinct mechanisms: absorption of the dopants by the 2D crystal or adsorption on its surface.
To distinguish the relevant mechanism, we systematically dope 53 experimentally synthesized 2D monolayers by 65 different chemical elements in both absorption and adsorption sites.
The resulting 17,598 doped monolayer structures were generated using the newly developed ASE \texttt{DefectBuilder}---a Python tool to set up point defects in 2D and bulk materials---and subsequently relaxed by an automated high-throughput density functional theory (DFT) workflow.
We find that interstitial positions are preferred for small dopants with partially filled valence electrons in host materials with large lattice parameters.
On the contrary, adatoms are favored for dopants with a low number of valence electrons due to lower coordination of adsorption sites compared to interstitials.
The relaxed structures, characterization parameters, defect formation energies, and magnetic moments (spins) are available in an open database to help advance our understanding of defects in 2D materials.

\end{abstract}

\maketitle
\ioptwocol

\section{Introduction}
Atomically thin 2D materials constitute promising material platform for building advanced nanoscale devices~\cite{ferrari2015science,schaibley2016valleytronics,sierra2021van,lin2019two} with unique control of electrons down to the level of individual quantum states~\cite{liu20192d,turiansky2020spinning}.
The physical properties of 2D materials can be tuned in a variety of ways, \textit{e.g.} by applying mechanical strain~\cite{dai2019strain,conley2013bandgap} or electric fields~\cite{yu2009tuning,leisgang2020giant,peimyoo2021electrical}, stacking monolayers into multilayers~\cite{novoselov20162d}, molecular functionalization \textit{via} their surface~\cite{brill2021molecular} or introducing dopants.
Although the introduction of dopants can have a detrimental impact on certain materials properties, such as carrier mobility or lifetime~\cite{polman2016photovoltaic}, they can also be used to control the amount of charge carriers in semiconductors or even instill new properties such as localized electron states with distinct emission lines~\cite{awschalom2013quantum,gomonay2018crystals,eckstein2013materials,gardas2018defects}, magnetism~\cite{friend1987electronic,zhao2020engineering,coelho2019room,wang2016robust}, or active catalytic sites~\cite{chen2017emerging,jia2018defect,tang2017nanocarbon,yan2017defect}. 

When impurity atoms dope a 2D material, the precise position of the dopants, in particular, whether they are located in the interior or on its surface, is decisive for how they influence the properties of the material. 
For example, for monolayer transition metal dichalcogenides (TMDs), it has been shown that the incorporation of metal dopants inside a 2D material can induce compositional phase changes~\cite{coelho2018post} whereas adsorption has a big impact on catalytic activity~\cite{wang2018design} or surface-enhanced Raman scattering~\cite{li2018ag}.
This makes it essential to establish the relative stability of adsorption \textit{versus} absorption sites for 2D dopants in general. 
Previous first principles studies have shown that 2D TMDs doped by transition metal atoms can favor either internal or surface dopant sites, depending on the dopant species~\cite{karthikeyan2019transition}.
However, while first principles calculations have been widely used to investigate the role of specific dopants in specific 2D host materials~\cite{karthikeyan2019transition,fu2017tuning,costa2020unveiling}, there exists to date no systematic study of doping in 2D materials across many different host materials, dopant sites, and dopant species. 

In this study, we turn to high-throughput calculations to answer whether a given dopant adsorbs---stays on the surface as an adatom---or absorbs---goes into the material as an interstitial---when used to dope a 2D material.
The process in focus is deposition of dopants on the monolayer.
Since substitutionals require the removal of an host atom, in other words a change in
stoichiometry compared to just the addition of an interstitial, they are omitted form this study.
We systematically dope 53 experimentally known 2D monolayers from the Computational 2D Materials Database (C2DB)~\cite{haastrup2018computational,gjerding2021recent} with 65 different atomic species in interstitial positions and adsorption sites (for further details on the data set, see Sec.~\ref{sec:host}).

To facilitate the structure set up, we implement the \texttt{DefectBuilder} module of the Atomic Simulation Environment (ASE)~\cite{larsen2017atomic} based on a defect generation scheme from the Automatic Defect Analysis and Qualification (ADAQ) software~\cite{davidsson2021adaq,adaq_info} originally
designed and used for bulk materials.
The \texttt{DefectBuilder} is extended to also include 2D materials.
With the DefectBuilder, 17,598 defect systems were created and later processed in an automatic workflow where 13,004 defect systems were fully relaxed and included in the analysis (cf. Sec.~\ref{subsec:workflow}).
For each host-dopant combination, we evaluate the formation energy of the dopant atom in a range of inequivalent interstitial and adsorption sites.
We analyze the preference for adsorption versus absorption, identify general trends in the data set of 13k relaxed defect structures, and collect the data in an open-access database, which should be useful as a resource for future investigations of impurity doping in 2D materials.

Our calculations use  DFT with the PBE exchange-correlation functional, which is known to have difficulties describing the localized states in, e.g., transition metals.
Nevertheless, the advantages of using the PBE functional for all systems and dopants are:
(i) consistency - data calculated on the same level of theory makes direct comparison of results easier.
Use of, e.g., DFT+U~\cite{Anisimov_1997} complicates comparisons of energies and leads to difficult decisions in regard to what U values to use;
(ii) benchmark - our results can be compared with other similar efforts that used the PBE functional~\cite{karthikeyan2019transition};
(iii) computational effort - the PBE functional
is computationally efficient, and the relaxed structures can be used as starting points for more accurate methods.

The paper is organized as follows: Sec.~\ref{sec:results} first introduces the set of host materials, explains the ASE \texttt{DefectBuilder} tool used to set up the initial structures, and defines key parameters for the interpretation of the results.
Afterward, the methodology is benchmarked against existing data in the literature for the specific class of 2H-MoX$_2$ monolayers, and subsequently, general trends in the entire data set are discussed.
Finally, we summarize our findings and look ahead in Sec.~\ref{sec:conclusion}.
The 'Methods' Section details the computational workflow and presents the resulting database.
The Supporting Information analyzes the numerical convergence and success rate of the high-throughput DFT calculations and finds clear trends that could be helpful as guidelines for future studies of similar nature.

\section{Results}\label{sec:results}

\subsection{Host materials}\label{sec:host}
The set of host materials was selected by screening the Computational 2D Materials Database (C2DB) ~\cite{haastrup2018computational,gjerding2021recent} for materials previously synthesized in monolayer form.
From the resulting 55 monolayers, we removed the one-atom-thick materials graphene and hexagonal boron nitride (hBN).
These materials were removed because:
(i) Absorption in interstitial sites is not well defined in such materials. 
(ii) Our calculations show that interstitials in fully planar systems are particularly challenging to converge with respect to in-plane supercell size (see Supplementary Note 1 of the Supporting Information).
(iii) The materials can exhibit a large variety of buckled structures depending on the dopant~\cite{lehtinen2015implantation}.
An overview of the host materials with their space group number is collected in Supplementary Note 2 of the Supporting Information.
For a selection of host materials, we performed convergence tests to determine the minimal supercell size needed for reliable results, see Supplementary Figure 1 of the Supporting Information. Based on these tests, supercells ensuring defect-defect distances of at least 10 {\AA} were chosen for all calculations.

\subsection{DefectBuilder}
\begin{figure*}
  \centering
  \includegraphics[width=\textwidth]{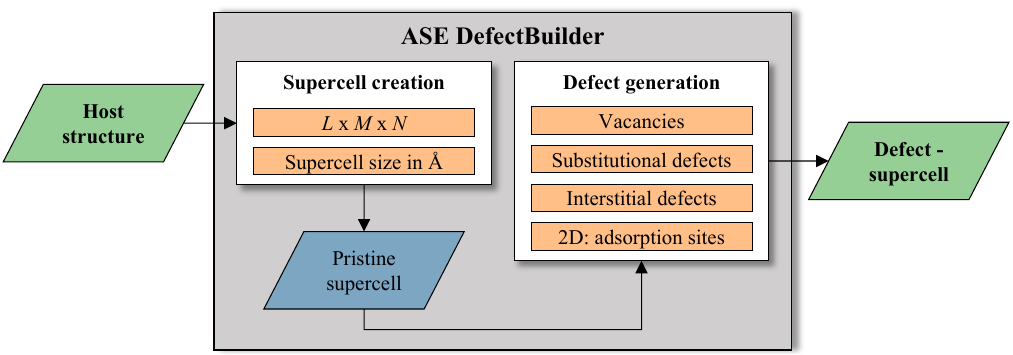}
  \caption{\textbf{Structure of the \texttt{DefectBuilder} class in ASE.} The \texttt{DefectBuilder} takes a host input structure in the commonly known formats used by ASE and sets up a pristine supercell (which can either be defined by simple \textit{L}x\textit{M}x\textit{N} repetitions or a physical minimum defect-defect distance criterion). Afterwards, the module generates different kinds of single point defects and finally returns the desired defect structures in their respective supercells.}
  \label{fig:2.1_defectbuilder.overview}
\end{figure*}
Large-scale studies of crystal point defects rely on tools to automatically define and set up the relevant defect structures.
In this work, we implement the ASE \texttt{DefectBuilder}, a useful module within the Atomic Simulation Environment~\cite{larsen2017atomic} to generate defect structures and supercells.
Figure~\ref{fig:2.1_defectbuilder.overview} gives an overview of the functionalities currently supported by the \texttt{DefectBuilder} which features simple functionalities to set up suitable defect supercells, \textit{e.g.}, by specification of a minimum distance between periodic repetitions of the defect.
After supplying the host crystal as an input structure in one of the numerous ASE structure formats, the \texttt{DefectBuilder} can generate single point defects like (i) vacancies, (ii) substitutional defects, (iii) interstitial defects, and (iv) adsorption sites for quasi-2D materials (including slabs used as a model for the surface of a bulk structure).

For (i) and (ii), the module analyzes the Wyckoff positions of the input structure and generates vacancies, antisites, or substitutionals (with selected elements) for each inequivalent position.
For (iii), the creation of interstitial defects, is based on the algorithm developed for the ADAQ framework~\cite{davidsson2021adaq}.
This algorithm produces a Voronoi tessellation of the host crystal.
The corners and centers of edges of the Voronoi cells are selected as the possible interstitial sites and a symmetry analysis discards equivalent sites.
One input determines the minimum distance between interstitial positions and atomic positions of the host crystal which controls the number of generated interstitial sites.
A larger minimum distance will produce fewer interstitial sites.

The interstitial site generation algorithm is further generalized to set up adsorption sites (iv): for a given quasi-2D input structure, the algorithm extracts the atoms from the topmost and lowermost part of the structure and separate Voronoi tessellations are performed for each of the two planar atomic layers. Possible adsorption sites are selected as the corners and edge centers of the 2D Voronoi cells obtained by restricting the 3D cells to the atomic plane.
Afterward, the 2D interstitial sites are translated out of the material to the point where the minimum distance between the adsorbate site and the closest atom of the 2D material equals the sum of the covalent radius of dopant and closest atom in the 2D material.
More details, such as all input parameters, can be found in the source code~\cite{DBsource} and the documentation of the \texttt{DefectBuilder} class~\cite{DBdocs}.

\subsection{Classification parameters}
We introduce three parameters to analyze the relaxed atomic structures and energetics of our calculations: $D\left[\mathrm{H,X}\right]$, $\mathrm{XF}\left[\mathrm{H,X}\right]$, and $\Delta\left[\mathrm{H,X}\right]$.
In order to classify the relaxed defect structures as adsorption or absorption configurations, we introduce a \emph{depth parameter}, $D$. For a given dopant X in a host material H, the depth parameter is defined by
\begin{equation}
\label{eq:depthparameter}
    D\left[\mathrm{H,X}\right] =  \frac{2z\left[\mathrm{X}\right] - \left(z_\mathrm{min}\left[\mathrm{H}\right] + z_\mathrm{max}\left[\mathrm{H}\right]\right)}{z_\mathrm{max}\left[\mathrm{H}\right] - z_\mathrm{min}\left[\mathrm{H}\right]}.
\end{equation}
Here, $z_\mathrm{min}\left[\mathrm{H}\right]$ ($z_\mathrm{max}\left[\mathrm{H}\right]$) is the lowermost (topmost) $z$-position in the pristine host structure and $z\left[\mathrm{X}\right]$ denotes the $z$-position of the dopant atom. Values $\abs{D} < 1$ correspond to absorption in an interstitial site while $\abs{D} \geq 1$ implies an adsorption site.
The sign of $D$ indicates whether the dopant sits above or below the center of the pristine monolayer, and the values $-1$ and $+1$ correspond to the dopant sitting exactly at the lower or upper boundary of the host crystal.

For some systems, the addition of a dopant into the structure can lead to large distortions.
Ideally, we would like the defect to only introduce small local changes, not an entire reorganization of the host crystal.
To quantify the amount of distortion introduced by the dopant atom, we introduce the \emph{expansion factor}, $\mathrm{XF}\left[\mathrm{H,X}\right]$, as
\begin{equation}
    \mathrm{XF}\left[\mathrm{H,X}\right] = \frac{d_\mathrm{rel}\left[\mathrm{H,X}\right]}{d_\mathrm{unrel}\left[\mathrm{H,X}\right]},
\end{equation}
where $d_\mathrm{unrel}$ ($d_\mathrm{rel}$) denotes the thickness of the monolayer plus dopant before (after) relaxation.
A large expansion factor, \textit{i.e.} $\mathrm{XF}\left[\mathrm{H,X}\right] > 2$, indicates an unphysically large restructuring of the monolayer.
This can, for example, happen when a large atom is introduced in a tight interstitial volume and leads to a disintegration of the monolayer during relaxation. Not unexpectedly, we find a strong correlation between large expansion factors and unconverged calculations (here defined as more than 20 relaxation steps).

Lastly, to analyze the adsorption and absorption energetics of a given host and dopant combination, we introduce the quantity, 
\begin{equation}\label{eq:delta}
    \Delta\left[\mathrm{H,X}\right] = E^{f\mathrm{,min}}_\mathrm{int}\left[\mathrm{H,X}\right] - E^{f\mathrm{,min}}_\mathrm{ads}\left[\mathrm{H,X}\right],
\end{equation}
where $E^{f\mathrm{,min}}_x\left[\mathrm{H,X}\right]$ ($x=\mathrm{ads, int}$) is the minimum formation energy of a dopant X in host crystal H either at an adsorption or interstitial site (as defined by the depth parameter in Eq. (\eqref{eq:depthparameter})).
We note that $\Delta\left[\mathrm{H,X}\right]$ is only defined if at least one adsorption and interstitial configuration has been converged for the given system.
A negative value of $\Delta\left[\mathrm{H,X}\right]$ indicates that the interstitial position is more energetically favorable than the adatom position, and \textit{vice versa} for positive values of the parameter.
Furthermore, $\Delta\left[\mathrm{H,X}\right]$ is independent of the chemical potential as opposed to the absolute formation energy, $E^f$ (which is also available in the database). For the calculation of $E^f$, the chemical potential is taken as the energy of the dopant atom in its standard state, see Supplementary Note 3 of the Supporting Information.

\subsection{Transition metal doping of 2H-MoX$_2$ monolayers}\label{subsec:tmdoping}
\begin{figure*}
  \centering
  \includegraphics{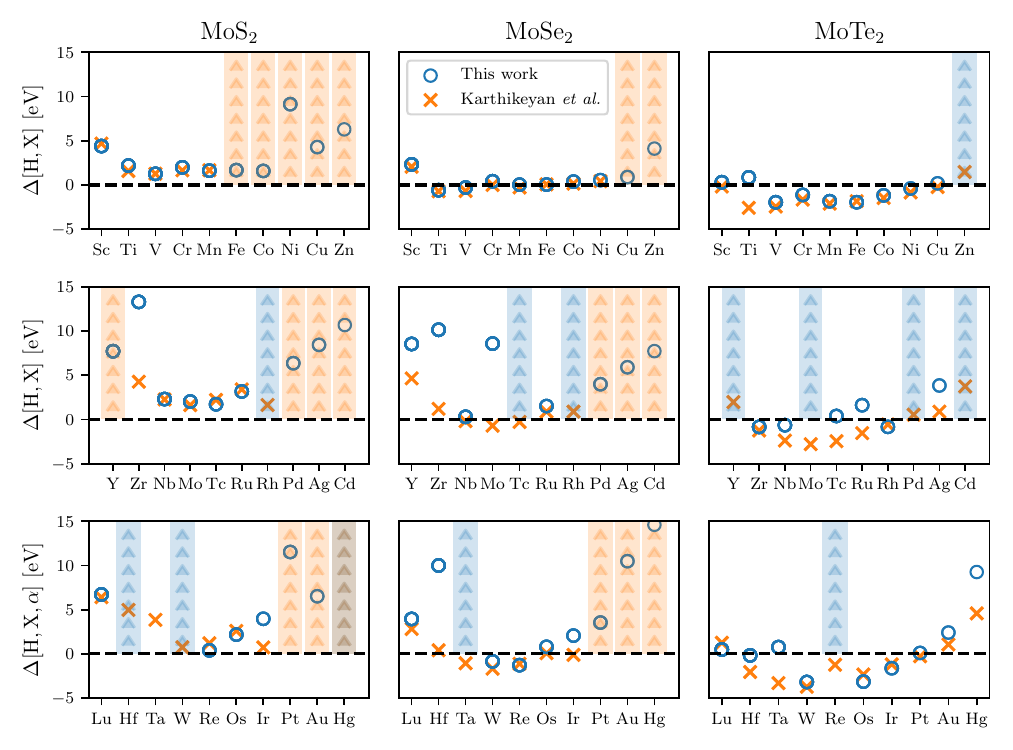}
  \caption{\textbf{Transition metal doping in transition metal dichalcogenides 2H-MoX$_2$.} $\Delta\left[\mathrm{H,X}\right]$ values for transition metal doped MoS$_2$ (left column), MoSe$_2$ (middle column) and MoTe$_2$ (right column) as a function of the doping element. The blue circles and blue bars show our calculated values whereas the orange crosses and orange bars are reference values for the same systems extracted from Ref.~\cite{karthikeyan2019transition}. The colored bars represent systems where not both adsorption sites and interstitial positions converged.}
  \label{fig:3.1_MoX2.doping}
\end{figure*}

Figure~\ref{fig:3.1_MoX2.doping} shows the $\Delta\left[\mathrm{H,X}\right]$ values (defined in Eq. (\eqref{eq:delta})) of transition metal-doped MoX$_2$ monolayers computed by our workflow.
Generally, the transition metal dopants are found to be more stable in adsorption sites (\textit{i.e.} $\Delta\left[\mathrm{H,X}\right]>0$) for MoS$_2$ and MoSe$_2$, whereas interstitial sites become more favorable (\textit{i.e.} $\Delta\left[\mathrm{H,X}\right]<0$) for MoTe$_2$.
This trend can be explained by the larger lattice constant of MoTe$_2$, which implies larger spaces to accommodate the dopant in an interstitial site.
This correlation is also well in line with our analysis of general convergence behavior, which is discussed in Supplementary Note 4 of the Supporting Information. 

Our results are in overall good agreement with the results from Karthikeyan \textit{et al.}~\cite{karthikeyan2019transition} apart from a few exceptions (indicated by orange or blue bars), namely: Zr and Ir in MoS$_2$; Zr, Mo, and Hf in MoSe$_2$; Ti, Tc, Ru, Ag, Ta, Hg in MoTe$_2$.
For these systems we (blue bars) or Karthikeyan \textit{et al.} (orange bars) obtain $\Delta\left[\mathrm{H,X}\right]$ values that are out of the scale.
Manual inspection of the systems show that the behavior is due to convergence problems for the relevant, lowest energy interstitial site.
The correct interstitial site has indeed been created by the \texttt{DefectBuilder}, but the DFT calculation did not converge, and thus, the data point was not included in the calculation of $\Delta\left[\mathrm{H,X}\right]$.
We further note that Karthikeyan \textit{et al.} used more accurate computational parameters (\textit{i.e.} 6x6x1 supercells and denser $k$-point sampling), explaining the small quantitative deviations (on the order of a few hundred meV) between their and our results.
Despite these disagreements, the benchmarking shows that our defect setup combined with the computational workflow yield reasonably accurate results and justifies the application of the methodology to the full data set of 53 host materials.
Supplementary Note 5 of the Supporting Information shows similar trends for WX$_2$ and NbX$_2$.

\subsection{General trends}
\begin{figure*}
    \centering
    \includegraphics{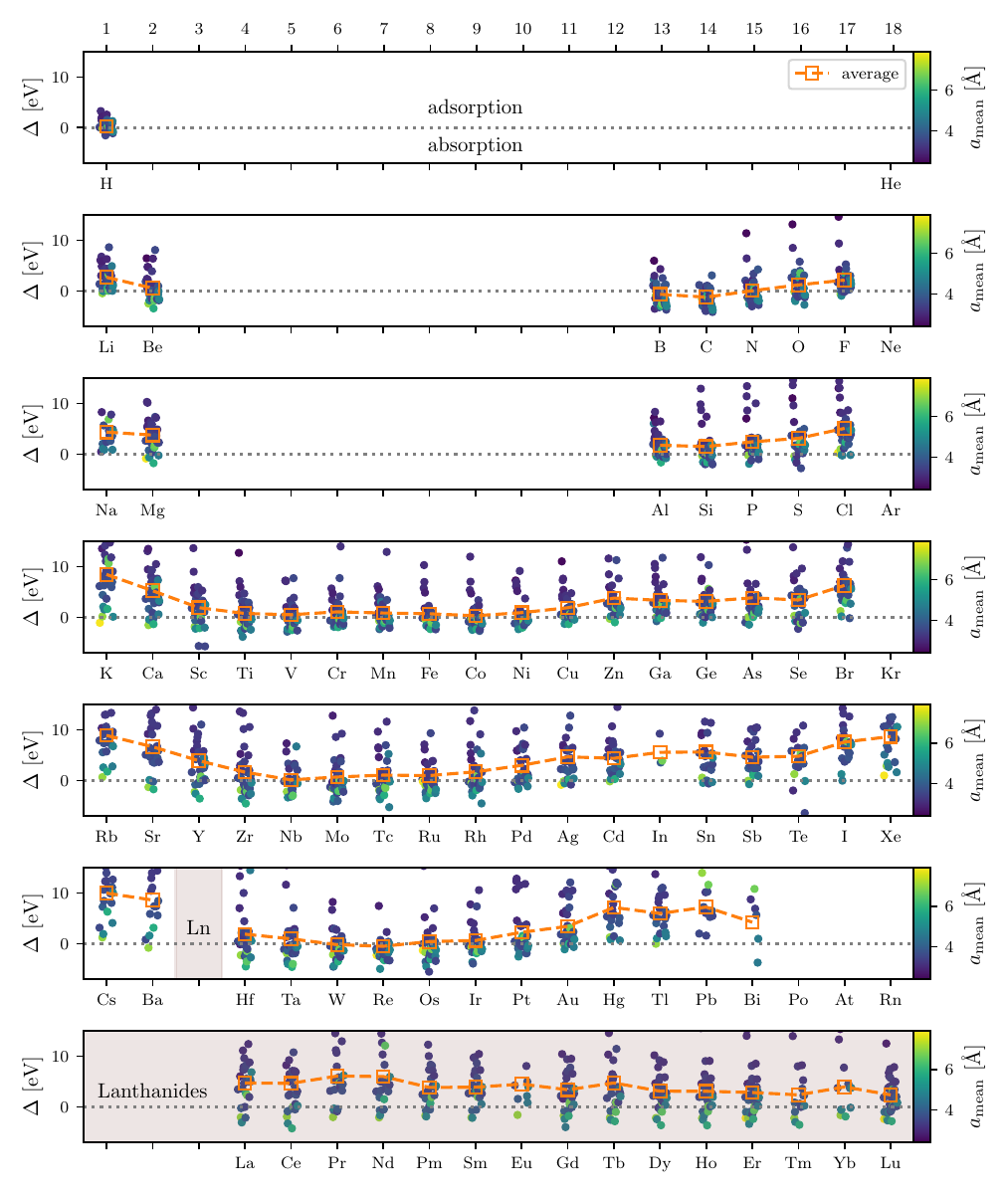}
    \caption{\textbf{Doping trends for all hosts and dopant elements.} $\Delta\left[\mathrm{H,X}\right]$ as a function of the respective dopant organized according to the periodic table order of the impurity species. Negative values correspond to stable interstitial. The different dots in one particular column represent the different host crystals. The average lateral size of the primitive host crystal unit cell is given by the color code of the data points. The orange squares visualize the average $\Delta$-value over all host crystals for a given dopant.}
    \label{fig:doping_trends}
\end{figure*}
After considering a few specific 2D monolayers and dopants in the previous section, we now explore trends in the entire data set of 53 host crystals and 65 dopants.
In particular, we focus on the question: which combinations of host material $\mathrm{H}$ and dopant atom $\mathrm{X}$ favor interstitial defects over adsorbates.
Figure~\ref{fig:doping_trends} shows the calculated $\Delta$-values for all the considered host crystals and doping elements.
The lattice constant (the average of the length of the in-plane basis vectors of the primitive unit cell) is indicated by the color code.

For dopants in the first row where only hydrogen was considered, the $\Delta$-values are distributed around zero, and there is no clear preference for adsorption or absorption.
For dopants of the second period, we see that Li and F prefer adsorption while B and C prefer absorption.
In contrast, for Be, N, and O the preference for adsorption/absorption is highly system dependent. 
Dopants from period 3 generally have larger $\Delta$-values, and most of the elements prefer adsorption. Exceptions occur for Si and, to a lesser extent, Al and P, which can also prefer absorption for specific materials.   

Dopants from periods 4-6 show very similar trends across the groups of the periodic table, indicating that the preference for adsorption/absorption is mainly dictated by the chemical nature of the dopant atoms.
Adsorption sites are favored for dopant elements from groups 1 and 2, whereas interstitial sites are preferred for the early and middle transition metal dopants with the exception of the group 3 elements (Sc, Y, Lu), which have a slight tendency to prefer adsorption.
We hypothesize that this effect can be explained by the interplay between the coordination number of a defect site and the number of available valence electrons for the dopant species.
On the one hand, adsorption sites possess a lower coordination number which is energetically favored by dopant species with a lower number of valence electrons, \textit{i.e.} groups 1, 2, and 3.
On the other hand, the coordination number of interstitial sites is generally higher due to more neighboring atoms inside the layer resulting in the preference of transition metals as dopant species.
In contrast, the late transition metals (group 10$-$12) generally favor adsorption due to a lack of valence electrons---the almost filled d-shell. The same holds for transition metals with a single d-electron.
Beyond the transition metal series, absorption is generally preferred. 
However, the $\Delta$-curve shows a convex shape as the $p$-shell fills. 
This is similar to the behavior observed for the transition metal series and supports the picture that absorption (adsorption) is generally favored when the dopant atom has more (fewer) valence electrons available for bonding.   


Not unexpectedly, there is a correlation between $\Delta$ and the lattice constant of the host crystal (indicated by the color coding in Figure~\ref{fig:doping_trends}): larger lattice constants are correlated with smaller $\Delta$-values.
This observation clearly indicates that the stability of interstitial sites is highly dependent on the available free space inside a monolayer and generalizes the corresponding trend observed in Sec.~\ref{subsec:tmdoping}.
Quantifying these correlations (\textit{e.g.}, by machine learning methods) appears worthwhile to explore in future studies.

Even if there are large variations in $\Delta$ depending on which host material the dopant is placed in, the general trend across all 2D host materials is clear: small dopants in spacious host materials are preferred.
The dopants of the s-block are large and rarely found as interstitials, except H which plays in a league of its own with an average $\Delta$ at zero with minimal variation across host materials.
The elements in Group 2 are smaller than in group 1, and the average $\Delta$ is lower for those elements.
Furthermore, the number of valance electrons also plays an important part.
Even if the elements in the p-block gets smaller as the group number increases, there are noticeable dips in the average $\Delta$ for the partially filled elements in Figure \ref{fig:doping_trends}.
Group 14 and 15 dopants have a lower average $\Delta$ than groups 13, 16, and 17.
This trend indicates that not just size is important but also the possibility to form bonds (see Supplementary Note 6 of the Supporting Information).
For the d-block, the elements do not vary noticeably in size and show large variations.
Also, the trend of partially filled valence is unclear from the average value.
Although, groups 3 and 12 have a higher average $\Delta$ compared to the rest.
For the sixth period, one can see that there are more points below the zero line.
Hence, the general trend is small dopants with partially filled valence electrons in spacious host materials are preferred.





\section{Discussion}\label{sec:conclusion}
We presented the ASE \texttt{DefectBuilder} -- a flexible and easy-to-use tool for setting up point defects and adsorption structures within the Atomic Simulation Environment (ASE)~\cite{larsen2017atomic}.
The ASE \texttt{DefectBuilder} is not limited to 2D materials and can be directly applied to study bulk systems and slabs. 
We utilized the \texttt{DefectBuilder} to systematically construct more than 17,500 interstitial point defects and adsorption structures by combining 65 dopant elements with 53 different 2D materials, which have all been experimentally realized in monolayer form~\cite{haastrup2018computational,gjerding2021recent}.
Each doped structure was subject to a relaxation and ground state workflow implemented within the \textit{httk}~\cite{httk,Armiento2020} high-throughput framework.

The computational approach was first benchmarked for transition metal-doped MoX$_2$ (X = S, Se, Te) monolayers and showed good agreement with previous studies~\cite{karthikeyan2019transition}.
In addition, interstitial and adsorption site stability trends in MoX$_2$ monolayers were generalized to other types of 2H-TMDs such as WX$_2$, and NbX$_2$.
Our results show that interstitial doping is generally very challenging to achieve over the entire set of 2D monolayers, especially for doping elements from the $s$- and $p$-blocks of the periodic table where the atoms are characterized by large covalent radii and/or few available electrons for bonding.
However, smaller elements like B, C, and N, as well as early to mid-transition metal atoms, are possible to introduce in interstitial sites of 2D materials that are not too closely packed. 

Looking ahead, data mining and machine learning techniques may be explored on the database to seek a more straightforward closed-form expression for predicting the configuration of an impurity atom. For example, a machine learning model can be trained to predict the formation energies of host materials outside the set considered here. 

In conclusion, all of the data produced in this work has been collected in an ASE database and is publicly available \textit{via} a web-application. 
This open-access approach can drive progress within single photon emission, transport applications, carrier lifetime evaluations, and other defect-mediated phenomena.
This database marks the starting point for future investigations of interstitial \textit{versus} adsorption site doping in 2D materials.

\section{Methods}\label{sec:methodology}
\subsection{Workflow}\label{subsec:workflow}
The calculations are carried out using the high-throughput toolkit (\textit{httk})~\cite{httk,Armiento2020} and the Vienna Ab initio Simulation Package (VASP)~\cite{VASP,VASP2}.
VASP implements density functional theory (DFT)~\cite{Hohenberg64,Kohn65} with the projector augmented wave (PAW)~\cite{PAW,Kresse99} method.
The Perdew, Burke, and Ernzerhof (PBE)~\cite{PBE} exchange-correlation functional is used, and all calculations are performed with spin polarization.
To speed up the calculations, the Brillouin zone (BZ) is sampled at the $\Gamma$-point only, which allows using the gamma compiled version of VASP for additional speed up.
Initial benchmarks performed for a subset of our systems show that the numerical error on defect formation energies due to the $\Gamma$-point approximation is below 100 meV.
The default VASP pseudopotentials~\cite{PAW_potentials} are used with the plane wave energy cutoff set to 600 eV and kinetic energy cutoff to 900 eV for all elements.
Calculations are performed for defects in their neutral charge state.

To ensure a fast and accurate relaxation of the vast number of defects, we employed a two-stage workflow inspired by ADAQ~\cite{davidsson2021adaq}.
The different settings for electronic and ionic tolerance as well as the Fast Fourier Transform (FFT) grid between the stages are shown in Table~\ref{tab:workflow}.
Both stages relax the atom positions and limit the ion relaxation to 20 steps.
Hence, a maximum of 40 ionic steps are taken for any given defect system.
The defect system does not have to reach the ionic tolerance in the final stage, the runs are saved to the database with the final ionic convergence. 
For the analysis in the main text, structures with a final ionic convergence of $5 \cdot 10^{-2}$ eV or less within 40 ionic steps are denoted as converged.
\begin{table}[h!]
\caption{\textbf{Settings for the automatic workflows.} The FFT grid column show the size compared to the largest wave vector.}
\centering
\begin{tabular} {c|ccc}
Stage & \makecell{Electronic\\tolerance [eV]} & \makecell{Ionic\\tolerance [eV]} & \makecell{FFT\\grid} \\
\hline
first & 10$^{-4}$ & $5 \cdot 10^{-3}$ & 3/2 \\
second & 10$^{-6}$ & $10^{-4}$ & 2\\
\end{tabular}
\label{tab:workflow}
\end{table}

\subsection{The database}
All of the interstitials and adsorption site systems have been subject to the workflow described in Sec. \ref{subsec:workflow}. As a result, we created more than 13,004 fully relaxed structures and collected them in an ASE database~\cite{larsen2017atomic}.
Each row of the database contains the relaxed atomic structure of the defect system and is uniquely defined by its host name (\texttt{host}), doping site (\texttt{site}, which can take the values 'int' and 'ads' followed by an internal integer index to distinguish between the different positions), and dopant atom (\texttt{dopant}).
Furthermore, we store numerous key-value pairs (KVPs) for easy querying of the data, \textit{e.g.} formation energy (\texttt{eform}), depth-parameter (\texttt{depth}), expansion factor (\texttt{expansion\_factor}), spin (\texttt{spin}), convergence (\texttt{converged}), \textit{etc.} 
Furthermore, a web application of the database will be available where users can interactively inspect the relaxed atomic structures of the interstitial and adsorption site doped materials, as well as all of their corresponding KVPs.
Finally, the database can be freely downloaded and accessed through its DOI (see 'Data availability' section).

\section{Acknowledgments}
The Center for Nanostructured Graphene (CNG) is sponsored by The Danish National Research Foundation (project DNRF103).
We acknowledge funding from the European Research Council (ERC) under the European Union’s Horizon 2020 research and innovation program Grant No. 773122 (LIMA) and Grant agreement No. 951786 (NOMAD CoE).
K. S. T. is a Villum Investigator supported by VILLUM FONDEN (grant no. 37789) and acknowledges funding from the Novo Nordisk Foundation Challenge Programme 2021: Smart nanomaterials for applications in life-science, BIOMAG Grant No. NNF21OC0066526.
R.A. and J.D. acknowledges funding from the Swedish eScience Centre (SeRC).
J.D. acknowledges support from the Swedish Research Council (VR) Grant No. 2022-00276.
R.A. acknowledges support from the Swedish Research Council (VR) Grant No. 2020-05402.
The computations were enabled by resources provided by the Swedish National Infrastructure for Computing (SNIC) at NSC and PDC partially funded by the Swedish Research Council through Grant Agreement No. 2018-05973.

\section{Data availability}
The data that support the findings of this study are openly available at the following URLs: \url{https://data.openmaterialsdb.se/imp2d/} and \url{https://doi.org/10.11583/DTU.19692238.v2}. The web-application of the database is available on the computational materials repository (CMR): \url{https://cmr.fysik.dtu.dk/imp2d/imp2d.html}.

\section{Code availability}
The sourcecode for the ASE \texttt{DefectBuilder} can be found on gitlab: \url{https://gitlab.com/ase/ase/-/tree/defect-setup-utils/ase}. Some simple code examples for the setup of defect structures is available at: \url{https://gitlab.com/ase/ase/-/blob/defect-setup-utils/doc/ase/build/defects.rst}.

\section{Competing interests}
The authors declare no competing interests.

\section{Author contributions}
J.D. and F.B. contributed equally.
J.D. and F.B. developed the initial concept, designed the code base, F.B. implemented the ASE \texttt{DefectBuilder} module, J.D. set up and developed the underlying workflow and ran the calculations, F.B. and J.D. analyzed the data and wrote the manuscript draft. K.S.T. and R.A. supervised the work and helped in interpretation of the results. All authors modified and discussed the paper together.

\section*{References}
\bibliographystyle{iopart-num}
\bibliography{references}

\end{document}


\newpage
\section{Supplementary Note 1: Supercell Size Convergence}
Before launching the study on the entire dataset, we conducted a convergence analysis with respect to in plane supercell sizes. The outcomes of this benchmark are shown in Supplementary Figure \ref{fig:sc_convergence}: in the left panel, we see the binding energy of an interstitial defects in MoS$_2$ and Bi$_2$I$_6$, respectively. One can see that the binding energy varies on the order of several hundreds of meV for the smallest supercell in MoS$_2$, but once the supercell sizes are larger than 10 {\AA} (\textit{i.e.} $\frac{1}{L} < 0.1$ {\AA}$^{-1}$), the binding energy values are well converged. Bi$_2$I$_6$, a system with more atoms in the primitive unit cell compared to MoS$_2$, converges for even larger supercell sizes. Based on these two observations we choose the minimum supercell size for the entire study to be at least 10 {\AA}. On the right hand side of Supplementary Figure \ref{fig:sc_convergence} we compare the convergence behaviour for an adatom and interstitial in the real planar host structure of hBN. The adatom is well converged already for small supercell sizes (which is intuitive as we do not expect adatoms to introduce big structural changes to the host crystal). The interstitial position, however, is not even converged for supecell sizes well above 10 {\AA}. Therefore, we excluded real planar systems like hBN and graphene from our full study.
\begin{figure}
    \centering
    \includegraphics{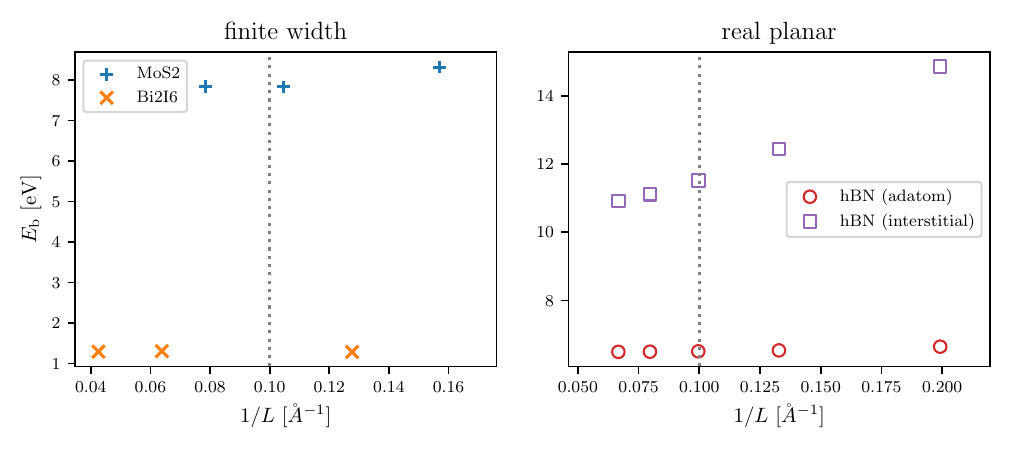}
    \caption{\textbf{Convergence behaviour of interstitial and adsorption sites in terms of supercell size.} Left: binding energy of an interstitial atom in MoS$_2$ and Bi$_2$I$_6$ (2D materials with finite out of plane width) as a function of the inverse supercell size $1\L$ where $L=\frac{L_x+L_y}{2}$ . Right: binding energy of an interstitial defect and atom in hBN (real planar system with no out of plane width) s a function of the inverse supercell size $\frac{1}{L}$.}
    \label{fig:sc_convergence}
\end{figure}

\newpage
\section{Supplementary Note 2: Overview of host materials}

\begin{longtable}{| c | c | c | c |}
    \caption{\textbf{List of host systems.} Overview of the 53 host crystals with host formula, space group, pristine band gap, magnetic information. Only Cr$_2$I$_6$ is a magnetic monolayer, all other host systems are non-magnetic.}
    \label{tab:hosts}
    \\
    \hline
    \textbf{Host formula}  & \textbf{Space group}  & \textbf{Band Gap}\\ 
    \hline
        P$_{4}$ & 53 & 0.90 eV \\
        Ga$_{2}$Se$_{2}$ & 123 & 0.00 e\\
        MoSe$_{2}$ & 187 & 1.32 eV \\
        W$_{2}$Te$_{4}$ & 11 & 0.00 eV \\
        ZrS$_{2}$ & 164 & 1.16 eV \\
        Bi$_{2}$I$_{6}$ & 147 & 1.38 eV \\
        Cr$_{2}$I$_{6}$ & 162 & 0.89 eV \\
        Hf$_{2}$Te$_{6}$ & 59 & 0.00 eV  \\
        PbI$_{2}$ & 164 & 1.50 eV  \\
        Nb$_{4}$C$_{3}$ & 164 & 0.00 eV \\
        NiSe$_{2}$ & 164 & 0.06 eV \\
        PtS$_{2}$ & 164 & 1.69 eV \\
        WTe$_{2}$ & 187 & 0.73 eV \\
        Si$_{2}$ & 164 & 0.00 eV \\
        Ti$_{3}$C$_{2}$H$_{2}$O$_{2}$ & 187 & 0.00 eV \\
        Ga$_{2}$Te$_{2}$ & 187 & 1.30 eV \\
        SnS$_{2}$ & 164 & 1.59 eV \\
        TaS$_{2}$ & 164 & 0.00 eV \\
        Mo$_{2}$CO$_{2}$ & 164 & 0.00 eV \\
        Sn$_{2}$ & 164 & 0.06 eV \\
        In$_{2}$Se$_{2}$ & 123 & 0.00 eV \\
        MoTe$_{2}$ & 187 & 0.93 eV \\
        NbSe$_{2}$ & 187 & 0.00 eV \\
        SnS$_{2}$ & 187 & 0.76 eV \\
        TiO$_{2}$ & 164 & 2.70 eV \\
        ZrSe$_{2}$ & 164 & 0.34 eV \\
        Ge$_{2}$H$_{2}$ & 164 & 0.92 eV \\
        Pd$_{2}$Se$_{4}$ & 14 & 1.31 eV \\
        SnSe$_{2}$ & 164 & 0.76 eV \\
        WS$_{2}$ & 187 & 1.53 eV \\
        Nb$_{2}$CO$_{2}$ & 164 & 0.00 eV \\
        GaN & 187 & 1.88 eV \\
        MoS$_{2}$ & 164 & 0.00 eV \\
        PtSe$_{2}$ & 164 & 1.17 eV \\
        Re$_{4}$S$_{8}$ & 2 & 1.27 eV \\
        MoS$_{2}$ & 187 & 1.58 eV \\
        BiITe & 156 & 0.41 eV \\
        As$_{2}$ & 164 & 1.48 eV \\
        C$_{2}$H$_{2}$ & 164 & 3.46 eV \\
        TiS$_{2}$ & 164 & 0.02 eV \\
        W$_{2}$Se$_{4}$ & 11 & 0.03 eV \\
        MoSSe & 156 & 1.45 eV \\
        HfS$_{2}$ & 164 & 1.22 eV \\
        TaSe$_{2}$ & 164 & 0.00 eV \\
        Ti$_{2}$CO$_{2}$ & 164 & 0.32 eV \\
        Re$_{4}$Se$_{8}$ & 2 & 1.12 eV \\
        WSe$_{2}$ & 187 & 1.24 eV \\
        V$_{2}$CO$_{2}$ & 164 & 0.00 eV \\
        C$_{6}$N$_{2}$ & 191 & 0.40 eV \\
        Ti$_{2}$S$_{6}$ & 11 & 0.29 eV \\
        Ge$_{2}$ & 164 & 0.03 eV \\
        HfSe$_{2}$ & 164 & 0.43 eV \\
        NbS$_{2}$ & 187 & 0.00 eV \\
        \hline
\end{longtable}
\newpage

\newpage
\section{Supplementary Note 3: Chemicial Potentials}

To evaluate the defect formation energy we set to the chemical potential of the dopant element to the energy of the crystalline phases~\cite{lejaeghere2014error}.
The elemental crystal structures were relaxed using a workflow similar to the one described in the main text, but with an initial volume relaxation step.
To ensure that the plane wave basis is accurate even when the volume is changed, the plane wave energy cutoff is increased to 700 eV and kinetic energy cutoff to 1400 eV.
The $k$-point grid is set up automatically with \textit{httk} using the Monkhorst-Pack method~\cite{monkhorst1976special}.
Electronic tolerance, ionic tolerance, and FFT grid are set to 10$^{-6}$ eV, $5 \cdot 10^{-5}$ eV, and twice the largest wavevector, respectively.
The last step is restarted until the energy difference between two steps is less than $5 \cdot 10^{-4}$ eV.
The formation energies are stored in the database.

\newpage
\section{Supplementary Note 4: Convergence statistics}\label{subsec:convergence}
Since a number of calculations did not converge within the set of ionic steps, it is of interest to check whether such systems could generally be assumed to be unphysical/unstable. For example, systems with a dopant placed in a small interstitial volume is unlikely to result in a stable configuration and is simultaneously expected to require many relaxation steps to reach a local minimum. If such a correlation exists, unconverged calculations would be less of a concern, as they would be unlikely to produce physically relevant information even if they were brought to convergence.
When considering all the calculations that did not converge in their ionic relaxation (in the following referred to as 'unconverged') we see that 36\% of the interstitial positions do not converge while only 20\% of the set up adsorption sites do not converge.

\begin{figure}
  \centering
  \includegraphics{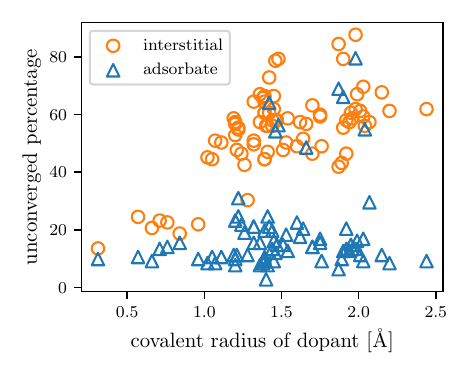}
  \caption{\textbf{Calculation of unconverged percentage and covalent radii.} Percentage of unconverged calculations as a function of covalent radius of the dopant element for adsorption sites (blue triangles) and interstitials (orange circles).}
  \label{fig:3.3_convergence.radius}
\end{figure}
\begin{figure}
  \centering
  \includegraphics{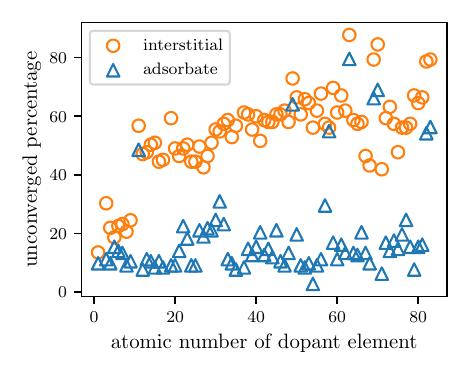}
  \caption{\textbf{Calculation of unconverged percentage and dopant atomic numbers.} Percentage of unconverged calculations as a function of atomic number of the dopant element for adsorption sites (blue triangles) and interstitials (orange circles).}
  \label{fig:3.3_convergence.atomic_number}
\end{figure}

Supplementary Figure \ref{fig:3.3_convergence.radius} shows the percentage of unconverged calculations as a function of the covalent radius of the dopant atom across all host crystals for interstitial and adsorption positions, respectively. It is clear that the rate of unconverged systems is higher for interstitial doping.
On one hand, the convergence rate for interstitials shows a clear correlation with the covalent radius of the dopant atom.
In general, it is difficult to converge calculations for interstitial dopants with covalent radius above $1$ {\AA}.
No such correlation is seen for dopants at adsorption sites, although there are a few outliers with above average unconverged rates.   

\begin{figure*}
    \centering
    \includegraphics{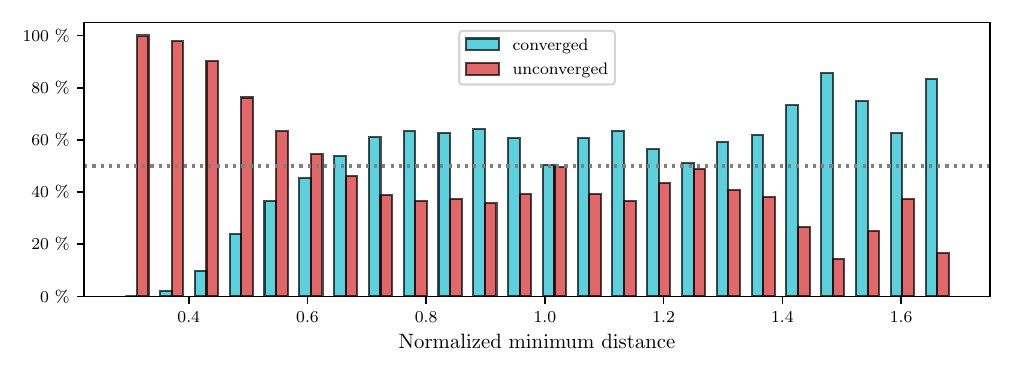}
    \caption{\textbf{Convergence behaviour and nearest neighbor distances for interstitial sites.} Bar diagram of the percentage of converged vs. unconverged calculations as a function of the normalized minimum distance parameter $D^\mathrm{norm}_{\mathrm{min}}\left(\mathrm{I}, \mathrm{A}\right)$. Each pair of bins adds up to 100 \%.}
    \label{fig:3.3_convergence.distances}
\end{figure*}

In Supplementary Figure \ref{fig:3.3_convergence.atomic_number}, a similar trend is obtained in terms of the unconverged percentage as a function of dopant type and atomic number of dopant element: the convergence for systems doped at adsorption sites is generally better and not strongly correlated with the atomic number of the dopant element.
For interstitials, however, a low unconverged rate is only present for dopant elements with atomic numbers lower than 15, whereas heavier dopant elements increase the percentage of unconverged calculations.

Interstitial sites seem to be the bottleneck, and an in-depth analysis of those systems is needed.
For a given system doped with an interstitial atom I, there exists one atom A in the host crystal that possesses the minimum distance $D_{\mathrm{min}}\left(\mathrm{I}, \mathrm{A}\right)$ to the interstitial site.
To quantify the convergence behavior of the interstitial systems, we define a normalized minimum distance parameter as
\begin{equation}
    D^\mathrm{norm}_{\mathrm{min}}\left(\mathrm{I}, \mathrm{A}\right) = \frac{D_{\mathrm{min}}\left(\mathrm{I}, \mathrm{A}\right)}{r_\mathrm{C}(\mathrm{I}) + r_\mathrm{C}(\mathrm{A})}.
\end{equation}
Here, $r_\mathrm{C}(\mathrm{A})$ and $r_\mathrm{C}(\mathrm{I})$ denote the covalent radii for the corresponding host atom and interstitial atom, respectively.
Supplementary Figure \ref{fig:3.3_convergence.distances} visualizes the percentage of converged and unconverged calculations for interstitial systems as a function of $D^\mathrm{norm}_{\mathrm{min}}\left(\mathrm{I}, \mathrm{A}\right)$.
For low normalized minimum distances almost no calculations converge but with increasing $D^\mathrm{norm}_{\mathrm{min}}\left(\mathrm{I}, \mathrm{A}\right)$ the percentage of converged calculations increases.
For normalized minimum distance values above 0.6, the probability of converging an interstitial position is generally larger than 50\%.
This trend is confirmed by the cumulative percentage of converged interstitial calculations: 56\% for $D^\mathrm{norm}_{\mathrm{min}}\left(\mathrm{I}, \mathrm{A}\right) > 0.6$ and only 25\% for $D^\mathrm{norm}_{\mathrm{min}}\left(\mathrm{I}, \mathrm{A}\right) < 0.6$. 

We note that the ionic convergence settings are set relatively low due to the vast number of systems that needed to be calculated.
In particular, not to waste computational resources, the ionic steps are limited to 20 since the ions may oscillate back and forth for some systems.
Some unconverged structures can be used as a good starting point for a detailed subsequent analysis. 
Unconverged systems can be found in the database \textit{via} the query: \texttt{converged=False}. 

In summary, we saw that our unconverged calculations can give insight into the stability of adsorption and interstitial site doping of 2D materials.
For interstitial defects, the overall large percentage of unconverged calculations stems from unphysical systems where an interstitial site simply does not have enough space inside the monolayer.
For instance, we saw that dopant atoms with large covalent radii are particularly challenging systems.
Therefore, in the context of high-throughput calculations, it is important to choose physically reasonable initial configurations for interstitial positions (we suggest normalized minimum distances of at least 0.6) to end up with convergence percentages of above 50\%.
On top of the structural challenges, our approach of doping the monolayers with most of the elements from the periodic table regardless of the chemical compatibility leads to large percentages of unconverged systems.
In a future study, a thorough analysis of the convergence behavior in terms of a chemical compatibility descriptor might be helpful for more insight.
The public availability of the dataset with its documentation online should make such an additional study easy to conduct.

The overall low percentage of converged structures is mostly due to the challenging task of relaxing interstitial defects. 
We note that a more computationally heavy workflow that allows for a larger number of ionic relaxation steps might improve that number in future studies.
Many of the generated initial structures turned out to be unphysical systems, and we suggest to always choose a reasonable normalized minimum distance when setting up interstitial structures.
Furthermore, the amount of unconverged calculations is correlated with the atomic species a given system is doped with, in particular for interstitials.

\newpage
\section{Supplementary Note 5: TM-doping for 2H-MX$_2$ Monolayers}
\begin{figure}
    \centering
    \includegraphics{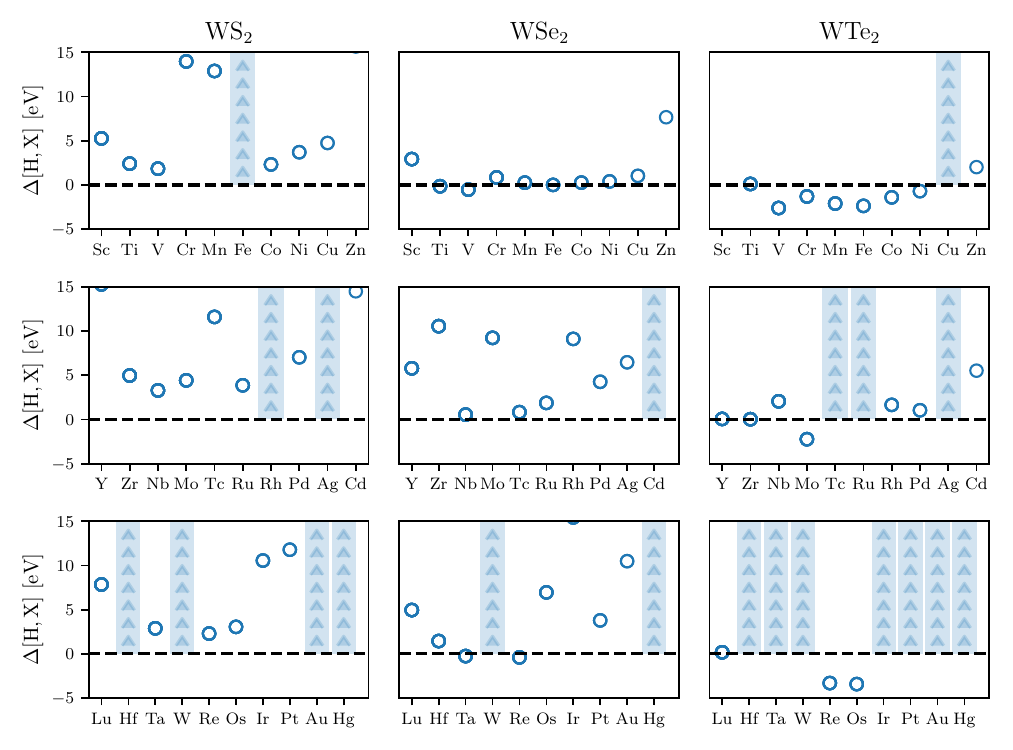}
    \caption{\textbf{Transition metal doping in transition metal dichalcogenides 2H-WX$_2$.} $\Delta\left[\mathrm{H,X}\right]$ values for transition metal doped WS$_2$ (left column), WSe$_2$ (middle column) and WTe$_2$ (right column) as a function of the doping element. Blue bars represent systems where we were not able to converge both adsorption sites and interstitial positions.}
    \label{fig:doping-WX2}
\end{figure}
\begin{figure}
    \centering
    \includegraphics{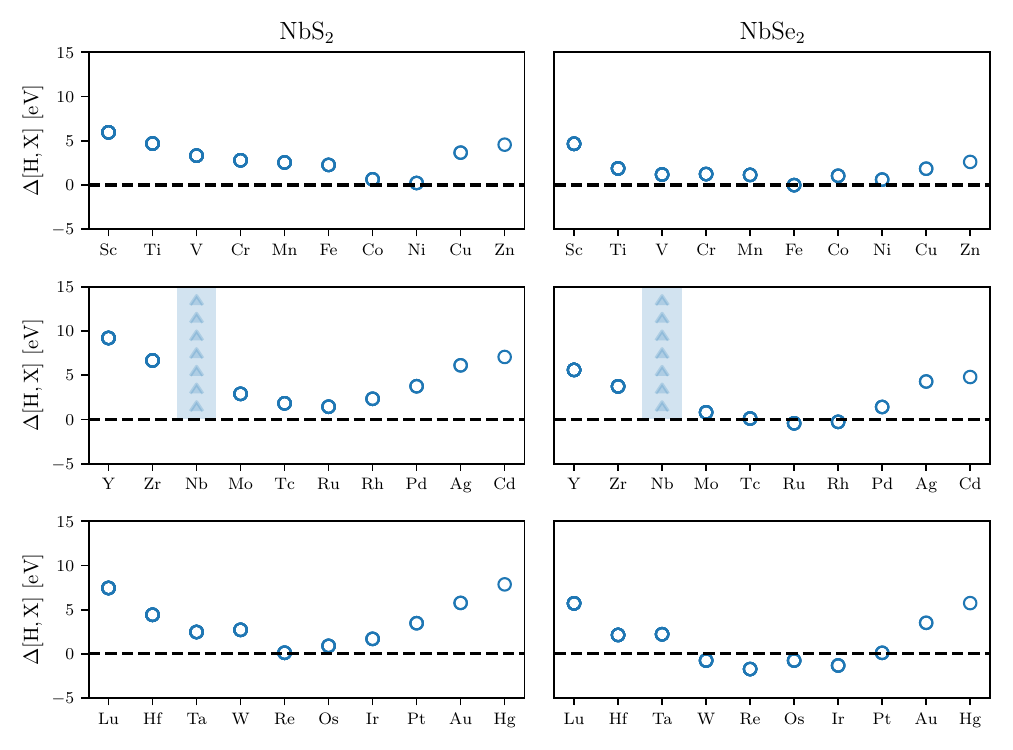}
    \caption{\textbf{Transition metal doping in transition metal dichalcogenides 2H-NbX$_2$.} $\Delta\left[\mathrm{H,X}\right]$ values for transition metal doped NbS$_2$ (left column), NbSe$_2$ (middle column) and NbTe$_2$ (right column) as a function of the doping element. Blue bars represent systems where we were not able to converge both adsorption sites and interstitial positions.}
    \label{fig:doping-NbX2}
\end{figure}

In addition to the interstitial and adsorption site trends observed for MoS$_2$, MoSe$_2$, MoTe$_2$, we conducted a similar analysis for the WX$_2$ (X=S,Se,Te) monolayers. The results are shown in Supplementary Figure \ref{fig:doping-WX2}: just like for the case of MoX$_2$ (see main manuscript), interstitial sites are not energetically favorable in the monolayer with the smallest in plane lattice constant, \textit{i.e.} WS$_2$. However, when going to the larger unit cells, interstitials become close in energy for WSe$_2$ and even favorable for the system with the largest lattice constant, \textit{i.e.} WTe$_2$. The same trend is also observed for NbS$_2$ and NbSe$_2$ (Supplementary Figure \ref{fig:doping-NbX2}).

\newpage

\bibliography{bibliography}